\documentclass[
,final            
  ]{aipproc}
\usepackage{epsfig}
\layoutstyle{6x9}
\begin{document}

\title {The brain: What is critical about it?}
\footnotetext{Proceeding of BIOCOMP2007 - Collective Dynamics:
Topics on Competition and Cooperation in the Biosciences.  Vietri
sul Mare, Italy (2007).}


\author{Dante R. Chialvo}
{address={Department of Physiology, Feinberg School of Medicine ,
Northwestern University, 303 East Chicago Ave. Chicago, IL 60611,
USA.}}
\author{Pablo Balenzuela}
{address={Departmento de F\'isica, Facultad de Ciencias Exactas y
Naturales, Universidad de Buenos Aires, Pabell\'on 1, Ciudad
Universitaria (1428), Buenos Aires, Argentina.}}
\author{Daniel Fraiman}
{address={Departamento de Matem\'atica y Ciencias,
   Universidad de San Andr\'es,  Vito Dumas 284 (1644), Buenos Aires,
   Argentina.}}

\begin{abstract}
We review the recent proposal that the most fascinating brain
properties are related to the fact that it always stays close to a
second order phase transition. In such conditions, the collective
of neuronal groups can reliably generate robust and flexible
behavior, because it is known that at the critical point there is
the largest abundance of metastable states to choose from. Here we
review the motivation, arguments and recent results, as well as
further implications of this view of the functioning brain.
\end{abstract}

\maketitle

\section{Introduction}

The brain is a complex adaptive nonlinear system that can be
studied along with other problems in nonlinear physics from a
dynamical standpoint. With this perspective here we discuss a
proposal
 \cite{bakbook,chialvo99,chialvo2004,chialvo2006,chialvo2007,chialvo2008}
claiming that the brain is spontaneously posed at the border of a
second order phase transition. The claim is that the most
fascinating properties of the brain are -simply- generic
properties found at this dynamical state, suggesting a different
angle to study how the brain works. From this viewpoint, all human
behaviors, including thoughts, undirected or goal oriented
actions, or simply any state of mind, are the outcome of a
dynamical system -the brain- at or near a critical state.

The starting point for this conjecture is that it is only at the
critical point that the largest behavioral repertoire can be
attained with the smallest number of degrees of freedom.
\emph{Behavioral repertoire} refers to the set of actions useful
for the survival of the brain and \emph{degrees of freedom} are
the number of (loosely defined) specialized brain areas engaged in
generating such actions. A number of ideas from statistical
physics can be used to understand how the brain works by looking
at the problem from this angle.

This article is dedicated to discussing the basis and specifics of
this proposition along with its implications. The paper is
organized as follows: The second section begins by reviewing the
problem. Basic features of the physics of critical phenomena are
introduced and used to support the Darwinian notion that brains
are needed to survive in a critical world. The third section
addresses predictable observations, and the fourth section reviews
recent results that support the idea of a critical state in brain
function. The paper closes with a short discussion of its
implications.

\section{What is the problem?}
New fascinating discoveries about brain physiology are reported
every week , each one uncovering a relatively isolated aspect of
brain dynamics. Yet the reverse process -how these isolated pieces
can be integrated to explain how the brain works- is rarely
discussed. Large-scale knowledge of the nervous system is
generally only casted in psychological terms, with little
discussion of underlying mechanisms. The goal should be, as it is
in physics, to explain all macroscopic phenomena -regardless of
their nature- on the basis of their underlying microscopic
dynamics.

The problem discussed here concerns  which  underlying properties
allow the brain to work as a collective of neuronal groups. How
chief brain abilities work in concert, how perception and action
are engaged, and how the conscious mind emerges out of electrical
impulses and neurochemistry is what we wish to understand, to name
a few. This is essentially equivalent, for instance, to
understanding how culture (or any other community emergent
property) emerges from each individual's intellectual capital. It
is clear that the solution of these questions, as for other
complex systems, requires more than the mere enumeration of all
the knowledge about the individual components.

The task of understanding how a collective works together is
challenging, but even more is in the case of the brain. As a whole
the brain has some notoriously conflictive demands. In some cases
it needs to stay ``integrated'' and in others must be able to work
``segregated'', as discussed extensively by Tononi and colleagues
\cite{Tononi98,Tononi2004}.\footnote{Perhaps the same conflict can
be identified also in other complex systems.} This is a non
trivial constraint, nevertheless mastered by the brain as it is
illustrated with plenty of neurobiological phenomenology. Any
conscious experience always comprises a single undecomposable
scene \cite{Tononi98}, i.e., an integrated state. This integration
is such that once a cognitive event is committed, there is a
refractory period in which nothing else can be thought of. At the
same time segregation properties allow for a large number of
conscious states to be accessed over a short time interval. As an
analogy, the integration property we are referring to could be
understood as the capacity to act (and react) on an all-or-nothing
basis, similar to an action potential or a travelling wave in a
excitable system. The segregation property could be then
visualized as the capacity to evoke equal or different
all-or-nothing events using different elements of the system. In
fact, this metaphor may be more than applicable.

It will be discussed below that the segregation-integration
conflict shares many similarities with the dynamics seen in
nonlinear systems near a order disorder phase transition.

\subsection{What is special about being critical}

Work in the last two decades has shown that complexity in nature
often originates from the tendency of non-equilibrium extended
nonlinear systems to drift towards a critical point. There are
many examples in which this connection was made more or less
rigourously including problems in physics, economy, biology,
macroeconomics, cosmology and so on \cite{Mayabak,
sole,turcotte,bakbook,buchanan}. It had been argued already
\cite{bakbook,chialvo99,chialvo2004,chialvo2007,chialvo2008} that
the same approach should be used to understand large scale brain
dynamics.

To review such a proposal we will briefly discuss which features
of the critical state are pertinent to the conjecture that the
brain is critical. As an example we will use the well known Ising
model of the magnetization in ferromagnetic materials, but it
should be noted that the important point are the universal
features of the phase transition and not the model itself.

We can describe the Ising model by considering a relatively small
square lattice containing $N=LxL$ sites, with each $i$ site
associated with a variable $s_i$, where $s_i=+1$ represents an
``up'' spin and $s_i=-1$ a ``down'' spin. Then any particular
configuration of the lattice is specified by the set of variables
${s_1,s_2,...,s_N}$. The energy in absence of external magnetic
field is given by
\begin{equation}
E = -J\sum_{i,j=nn(i)}^N s_is_j
\end{equation}
where  $J$ is the coupling constant and the sum of $j$ runs over
the nearest neighbors of a given site $i$ ($nn(i)$). The
simulation is usually implemented with the Metropolis Monte Carlo
algorithm \cite{Metro, Tobo} solving for a given heat bath
temperature $T$.

Collectively, spins will show different degrees of order and
magnetization values depending on the temperature, as seen in the
ferromagnetic-paramagnetic phase transition illustrated in Figure
1. A material is ferromagnetic if it displays a spontaneous
magnetization in absence of any external magnetic field. If we
increase the temperature the magnetization gets smaller and
finally reaches zero. At low temperature the system is very
ordered with only very large domains of equally oriented spins, a
state almost invariant in time.  At very high temperatures, spin
orientation changes constantly and become correlated only at very
short distances resulting in vanishing magnetization. Only in
between these two homogeneous states, at the critical temperature,
does the system exhibit peculiar fluctuations both in time and
space. The temporal fluctuations of the magnetization is scale
invariant. Similarly, the spatial distribution of spins clusters
show long range (power law) correlations and scale invariance
reflected in a fractal structure of clusters of aligned spins. It
is important to realize two points: 1) these large structures only
emerge at the critical point, and 2) they extend up to the system
size despite the fact that the interactions between the systems
elements are only \emph{short-range} (i.e., between the nearest
neighbors). Thus, at the critical temperature, the system is able
to maintain correlation between far away sites (up to the size of
the system) staying long periods of time in a given meta-stable
state but also exploring a large diversity of such states. This
behavior is reflected in the maximization of the fluctuations of
magnetization, a typical signature of a second order phase
transition.

We propose that this dynamical scenario -generic for any second
order phase transition- is strikingly similar to the
integrated-segregated dilemma discussed above, and is necessary
for the brain to operate as a conscious device. It is important to
note that there are no other conceivable dynamical scenarios or
robust attractors known to exhibit these two properties
simultaneously. Of course, any system could trivially achieve
integration and long range correlations in space by increasing
links' strength among faraway sites, but these strong bonds
prevent any segregated state.

\begin{figure}
\centering
\includegraphics[height=100mm]{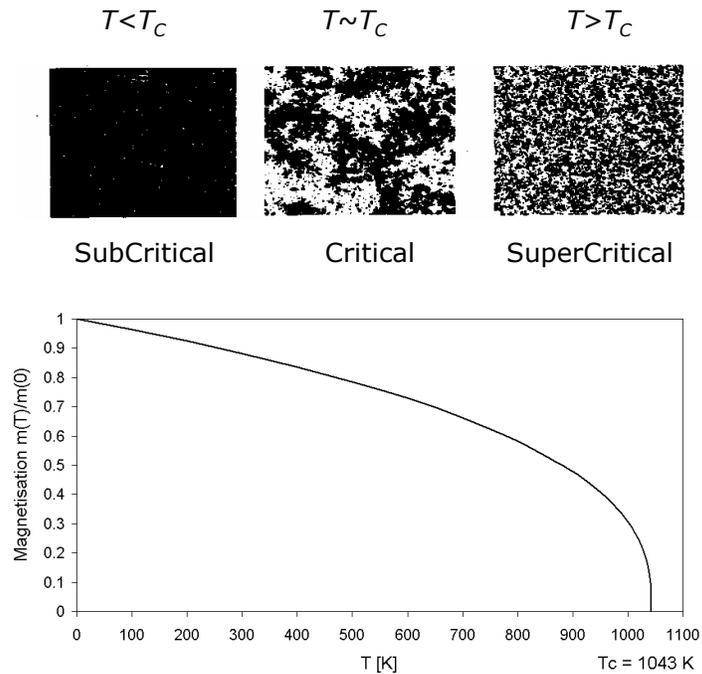}
\caption{Ferromagnetic-paramagnetic phase transition. Bottom:
Temperature dependence of magnetization m(T) for Fe. Top three
panels are snapshots of the spins configuration at one moment in
time for three temperatures: subcritical, critical, and
supercritical.}
\end{figure}

\subsection{Why do we need a brain?}
This question may sound frivolous but it is not at all, because in
Darwinian terms it is necessary to consider the brain embedded in
the rest of nature, and co-evolving according with the constraints
of natural selection. Although some views could advocate for
computational properties in specific neural circuits and find
mathematical justification for it existence, we simply think that
the brains we see today are the ones that -for whatever means- got
an edge and survived. How consistent is our view of the brain near
a critical point will be answered by considering these Darwinian
constraints. We propose that the brains we see today are critical
because the world in which they have to survive is up to some
degree critical as well. Let us look at the other possibilities.
If the world were sub-critical then everything around will be
simple and uniform (as in the left top panel of Figure 1); there
would be nothing to learn, a brain will be superfluous. In a
supercritical world, everything would be changing all the time (as
in the right top panel of Figure 1); in these conditions it would
be impossible to learn. Thus in neither extreme could a brain have
provided an edge to survive- in the very uniform world there is
nothing to learn and in the wildly fluctuating one there is no use
for learning. The brain, therefore, must only be necessary to
navigation in a complex, critical world.\footnote{It has been
already argued elsewhere \cite{bakbook,Mayabak} that the world at
large is critical.} In a critical world, things are most often the
same, but there is always room for surprises. To us, this is
-intuitively speaking- how the dynamics with power law
correlations look like, there is always a very unlikely event that
always surprises us, i.e., some novelty on a background of well
known usual things. We ``need'' a brain \emph{because} the world
is critical \cite{bakbook,bak1,bak2,chialvo99,Mayabak}.

Furthermore, a brain not only needs to learn and remember, but
also has to be able to forget and adapt. If the brain were
sub-critical then all brain states would be strongly correlated
with the consequence that brain memories would be frozen. On the
other extreme, a supercritical brain would have patterns changing
all the time, resulting in the inability to hold any long term
memory. One must conclude therefore that in order to be highly
susceptible the brain itself has to be near the critical state.

Of course these ideas are not entirely new, indeed almost the same
intuition prompted Turing half a century ago to speculate about
learning machines using similar terms \cite{Turing}.

\section{What should be seen?}

In previous writings we have advanced a tentative list of features
of the critical point that should be observed in brain
experiments. These included:

\begin{enumerate}

\item At large scale:\\
 Cortical long range correlations in space and time.
 Large scale anti-correlated cortical states.
 \item At small scale:\\
``Neuronal avalanches'', as the normal homeostatic state for most
neocortical circuits.
 ``Cortical-quakes'' continuously shaping the large scale synaptic landscape providing ``stability'' to the cortex.
\item At behavioral level:\\
 All adaptive behavior should be ``bursty'' and
apparently unstable, always at the ``edge of failing''.
 Life-long learning should be critical due to the effect of continuously ``rising the bar''.
\end{enumerate}

In addition one should be able to demonstrate that a brain
behaving in a critical world performs optimally at some critical
point, thus confirming the intuition that the problem can be
better understood by considering the environment from which brains
evolved.

In the list above, the first item concerns the most elemental
facts about critical phenomena: despite the well known short range
connectivity of the cortical columns, long range structures appear
and disappear continuously. The presence of inhibition as well as
excitation together with elementary stability constraints
determine that cortical dynamics should exhibit large scale
anti-correlated structures as well \cite{fox_05}. The features at
smaller scales could have been anticipated from theoretical
considerations, but avalanches were first observed empirically in
cortical cultures and slices by Plenz and colleagues
\cite{Plenz03}.  An important point that is left to understand is
how these quakes of activity shape the neuronal synaptic profile
during development. At the next level this proposal suggests that
human (and animal \cite{boyer,ramos}) behavior itself should show
evidence of criticality and learning also should be included. For
example, in teaching any skill one chooses increasing challenging
levels that are easy enough to engage the pupils but difficult
enough not to bore them. This ``raising the bar'' effect continues
through life, pushing the learner continuously to the edge of
failure! It would be interesting to measure some order parameter
for sport performance to see if shows some of these features for
the most efficient teaching strategies.

\section{Recent Results}
\subsection{Neuronal avalanches in cortical networks}

The first demonstration that neuronal populations can exhibit
critical dynamics were the experiments reported by Plenz' lab
\cite{Plenz03}. What they uncovered was a novel type of electrical
activity for the brain cortex. This type of population activity,
which they termed ``neuronal avalanches'', sits half way in
between two well known patterns: the oscillatory or wave-like
highly coherent activity on one side and the asynchronic and
incoherent spiking on the other. In each neuronal avalanche it is
typical of a large probability to engage only few neurons and a
very low probability to spread and activate the whole cortical
tissue. In very elegant experiments Plenz and colleagues estimated
a number of properties indicative of critical behavior including a
power law with an exponent $\sim 3/2$ for the density of avalanche
sizes (see Figure 2). This agrees exactly with the theoretical
expectation for a critical branching process \cite{Zapperi1995}.
Further experiments in other settings, including monkey and rat in
vivo recordings, have already confirmed and expanded upon these
initial estimations \cite{Plenz04,PlenzTINS, Stewart2006,Plenz07}.
\begin{figure}
\centering
\includegraphics[height=70mm]{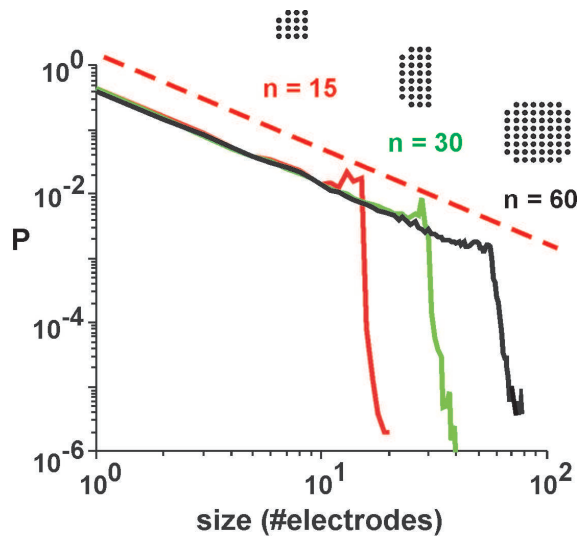}
\caption{Scaling in neuronal avalanches of mature cortical
cultured networks. The distribution of sizes follows a power law
with an exponent $\sim 3/2$ (dashed line) up to a cutoff which
depends on the grid size. The data, re-plotted from Figure 4 of
\cite{Plenz03}, shows the probability of observing an avalanche
covering a given number of electrodes for three sets of grid sizes
shown in the insets with n=15, 30 or 60 sensing electrodes
(equally spaced at $200 \mu m$). The statistics is taken from data
collected from 7 cultures in recordings lasting a total of 70
hours and accumulating 58000 ($+-$ 55000) avalanches per hour
(mean $+-$ SD).}
\end{figure}

An unsolved problem here is to elucidate the precise neuronal
mechanisms leading to this behavior. Avalanches of activity such
as the one observed by Plenz  could be the reflection of
completely different scenarios. It could be that the power law
distribution of avalanches sizes reflect several non- homogeneous
Poisson processes that when added together look like a scale free
process. This is unlikely, and scaling analysis should show that
this is not the case. It could also reflect a structural (i.e.,
anatomical) substrate over which travelling waves in the peculiar
form of avalanches occur. This would imply that the long range
correlations detected are trivially due to long range connections.
If that is the case, as was discussed above, this would have
nothing to do with criticality, and furthermore would imply that
segregation is impossible. Based of what is known about the
connectivity, it is reasonable to think of a dynamical mechanism
responsible for this type of activity. One can assume that the
neuronal avalanches occur over a population of locally connected
neurons. Their ongoing collective history will permanently keep
them near the border of avalanching and each collective event will
only excite enough neurons to dissipate the excess of activity.
This is the most likely scenario, following the ideas put forward
by Bak and colleagues \cite{bakbook,bak1,bak2,Mayabak}; however,
 there is no theoretical formalization of these results as of yet.

The most significant theoretical effort to elucidate the
mechanisms underlying neuronal avalanches was reported recently by
Levina and colleagues \cite{levina2007}. They considered a network
model of excitable elements with random connectivity in which the
coupling is activity dependent, such that, as in reality, too much
activity exhausts the synaptic resources. This induces a
decreasing in coupling strength which in turn decreases the
propagation of activity. The interaction between activity and
coupling results in a self-organized drifting of the dynamic
towards a critical avalanching activity with the statistics
reported in Plenz' experiments. Further work is needed to see
other spatiotemporal properties of neuronal avalanches to check if
they follows the mechanism suggested by Levina et al.
\cite{levina2007}.

\subsection{Functional brain networks are complex}
Functional magnetic resonance imaging (fMRI) allows us to
non-invasively monitor spatio-temporal brain activity under
various cognitive conditions. Recent work using this imaging
technique demonstrated complex functional networks of correlated
dynamics responding to the traffic between regions, during
behavior or even at rest (see methods in \cite{Eguiluz}. The data
was analyzed in the context of complex networks (for a review see
\cite{Sporns2004}). During any given task the networks were
constructed first by calculating linear correlations between the
time series of the blood oxygenated level dependent (BOLD) signal
in each of $36\times64\times64$ brain sites called voxels. After
that, links were defined between those brain sites whose BOLD
temporal evolutions were correlated beyond a pre-established value
$r_c$.
\begin{figure}
\centering
\includegraphics[height=120mm]{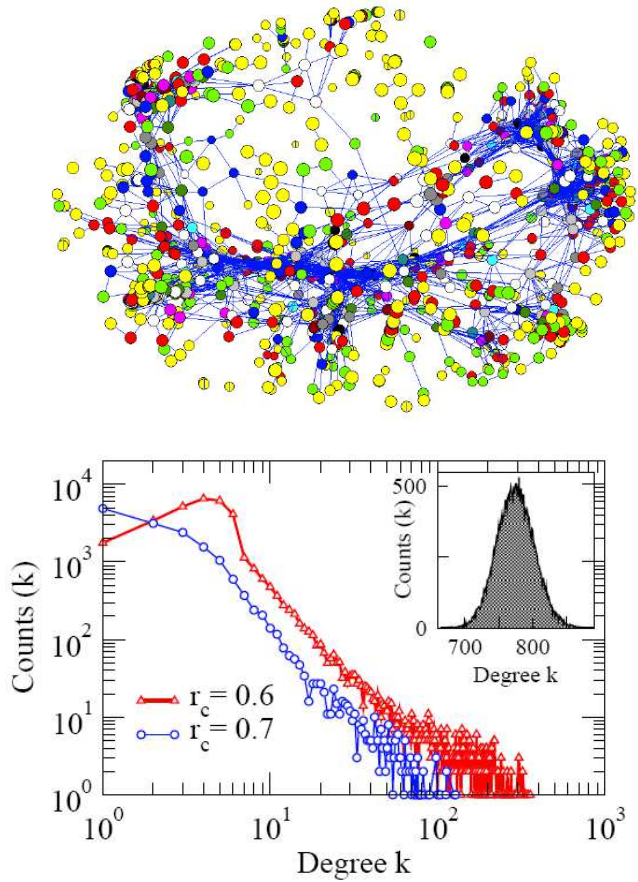}
\caption{A typical brain network extracted from the correlations
of functional magnetic resonance images. Top panel shows a
pictorial representation of the network. The bottom panel shows
the degree distribution for two correlation thresholds $r_c$. The
inset depicts the degree distribution for an equivalent randomly
connected network. Data re-plotted from \cite{Eguiluz}.}
\end{figure}

Figure ~3, show a typical brain functional network extracted with
this technique. The top panel illustrates the interconnected
networks' nodes and the bottom panel shows the statistics of the
number of links (i.e., the degree) per node.   There are a few
very well connected nodes in one extreme and a great number of
nodes with a single connection.  The typical degree distribution
approaches a power law distribution with an exponent around 2.
Other measures revealed that the number of links as a function of
-physical- distance between brain sites also decays as a power
law, something already confirmed by others \cite{Salvador} using
different techniques. Two statistical properties of these
networks, path length and clustering, were computed as well. The
path length ($L$) between two voxels is the minimum number of
links necessary to connect both voxels. Clustering ($C$) is the
fraction of connections between the topological neighbors of a
voxel with respect to the maximum possible.  Measurements of $L$
and $C$ were also made in a randomized version of the brain
network. $L$ remained relatively constant in both cases while $C$
in the random case were much smaller, implying that brain networks
are ``small world'' nets, a property with several implications in
terms of cortical connectivity, as discussed further in
\cite{Sporns2004b,Sporns2004,basset2006b,stamreview}. In summary,
the work in \cite{Eguiluz} shows that functional brain networks
exhibit highly heterogeneous scale free functional connectivity
with small world properties. Although these results admit a few
other interpretations, the long range correlations demonstrated in
these experiments are consistent with the picture of the brain
operating near a critical point, as will be further discussed
below. Of course, further experiments are needed to specifically
define and measure some order parameter to clarify the precise
nature of these correlations. Furthermore, as more detailed
knowledge of the properties of these networks is achieved, the
need to integrate this data in a cohesive picture grows
\cite{Sporns2006}.

To gain insight into the possible dynamical origins of Eguiluz
\cite{Eguiluz} findings we simulated the Ising model on a
relatively small square lattice at the critical temperature. Then,
as was done with the brain fMRI data, the linear correlations
between the time series of each one the lattice points
($s_i=\pm1$) were calculated:
\begin{equation}
r(i,j)= \frac{\langle s_i(t)s_j(t)\rangle -\langle s_i(t)\rangle
\langle s_j(t)\rangle}{\sigma(s_i(t)\sigma(s_j(t)))},
\end{equation}

where $\sigma^2(s(t)=\langle s^2(t) \rangle -\langle
s(t)\rangle^2)$.

Figure 4 illustrates typical results for the critical temperature.
The distribution of correlations is approximately Gaussian,
encompassing both positive as well as negative correlations (see
the left panel of Figure 4). This is related  to the large domains
of equally oriented spins found at the critical temperature, which
are positively correlated amongst themselves and negatively
correlated to domains with opposite spin orientation. These
counterbalanced correlations are only present at the critical
temperature, since for supercritical temperatures all correlations
vanish and for subcritical values only a large domain of a given
orientation survive.

In analogy with Eguiluz et al. methods, a correlation network was
constructed by defining links between those lattice points whose
fluctuations correlated beyond a a given $r_c$ value. The degree
distribution for $r_c=0.4$ is depicted in Figure 4 and 5, where it
can be seen that there is a mode centered around four (i.e., the
number of neighbors in the simulation) and then a long tail which
resembles very much the experimental results shown previously.
Further details can be appreciated more clearly in Figure 5. The
top right panel shows the degree for each lattice point, and the
top left a correlation map. Notice that the tail of the degree
distribution in the previous figure corresponds here to the points
in the two clusters with highest degree (colored yellow-red). In
the left panel, the origin of these clusters is clarified by
selecting one of them as a seed (labelled S) and plotting its
correlation values with the rest of the lattice points. Typical
time series of two nodes placed far away from the seed: one
positively correlated (P) and the other negatively correlated (N)
are also plotted in Fig. 5. Note that the two large
anti-correlated domains correspond to the two hubs in the degree
map.

Of course, these numerical experiments are very far from
representing anything close to the details of brain physiology.
Nevertheless, they serve the purpose of showing that key features
of the correlations seen in the fMRI experiments are also observed
in a paradigmatic critical system. The main point of these results
is to demonstrate that a correlation network with scale free
degree distribution as reported by Eguiluz et al. \cite{Eguiluz}
can be extracted from a dynamical system, providing is at a
critical point, regardless of the underlying connectivity. The
example shown here uses the worst case scenario of a lattice with
only \emph{local} connectivity, but we expect the main conclusions
to remain the same using other less ordered topologies.

It is important to remark that the dynamics described arises in
the Ising model with ferromagnetic interactions, i.e., there is
only positive correlations between neighbor sites (analogous to
have only ``excitatory synapsis''). Despite its absence in Eq. 1,
negative correlations emerge as a collective property of the
critical dynamics. Accordingly, these negative correlations
manifest at relatively long time scales (reflecting the collective
movements of spins) and not at short time scales. This agrees well
both with observations made from fMRI experiments and with those
extracted from a detailed model of the cortex \cite{honey}.
Finally another aspect to note is that the ratio between the area
covered by positive and negative correlations equal to one (see
Fig. 5), just as it observed in the brain of healthy people
\cite{baliki2} as discussed in the next section.

\begin{figure}
\centering
\includegraphics[width=0.85 \textwidth]{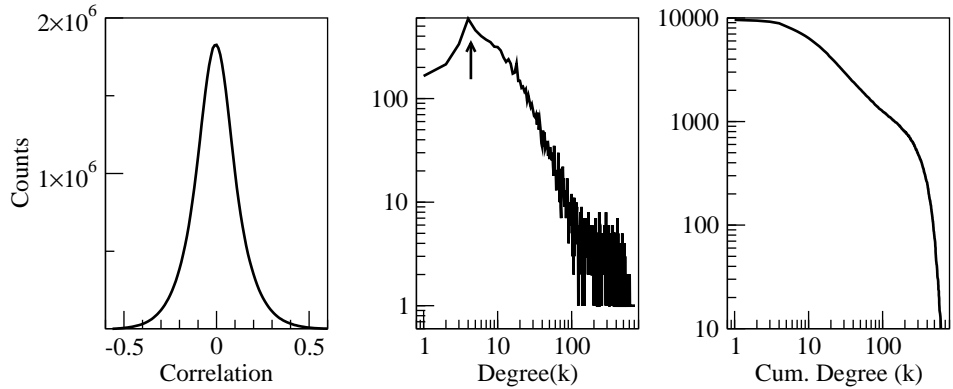}
\caption{Ising model at the critical temperature: Left plot shows
the distribution of correlation values. Middle and right plots
depict the degree distribution of the network extracted. Arrow
point at degree=nn=4. Correlation network constructed as in
\cite{Eguiluz} using a threshold $r_c > 0.4$ . Simulation of Eq. 1
with $k=1$ and $J=1$, discarding a transient $N_{equil}=10^8$
steps, we chose $N_{time}=1000$ configurations every
$N_{sample}=LxL=10^4$ steps. Each time step corresponds to a
single spin flip. In all cases the system is at the critical
temperature ($T_c \simeq 2.3J/k_B$)}
\end{figure}
\begin{figure}
\centering
\includegraphics[width=0.95\textwidth,angle=-90]{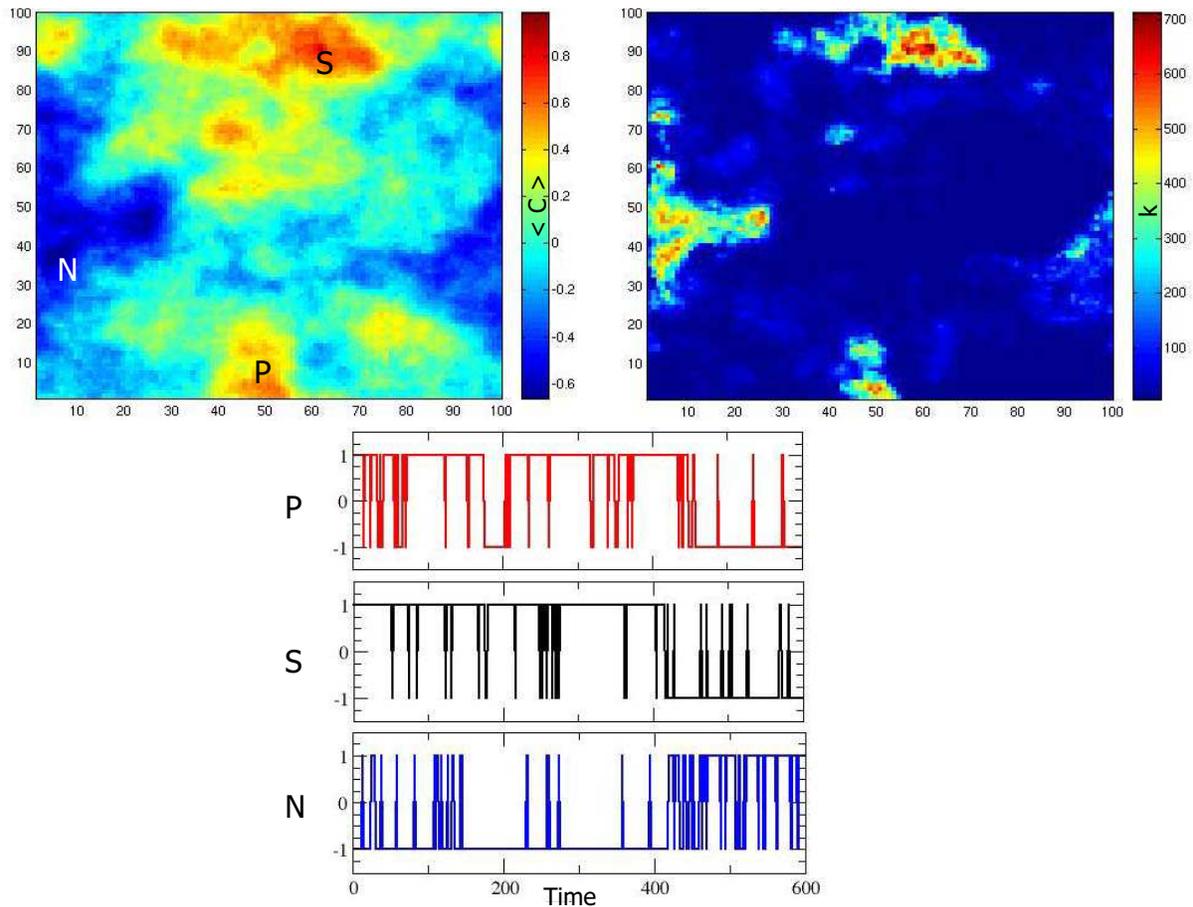}
\caption{Ising model at the critical temperature: Top left:
Correlations between the seed (S) and the rest of the lattice. Top
right: Degree (k) of each lattice point. Bottom: Time series of
three selected places, one for the seed (S) and one for a
negatively (N) or positively (P) correlated point. Each time step
here corresponds to $10^4$ single flip spins.}
\end{figure}

\subsection{What state is the brain ``resting state''?}

Over a decade ago \cite{Biswal} BOLD low-frequency fluctuations
were shown to be correlated across widely spatially separated but
functionally related brain regions (between left and right
sensorimotor cortices) in subjects at rest. Brain ``rest'' can be
defined - more or less unsuccessfully- as the state in which there
is no explicit brain input or output.\footnote{Readers familiar
with Italian traditions advantageously can specify brain rest as
the brain state resulting from \emph{``dolce fare niente''}.
Translated literally it reads ``sweet do nothing'' or also the
``sweet act of doing nothing''. }

Various groups have suggested that these fluctuations are of
neuronal origin and correspond to the neuronal baseline or idle
activity of the brain.  These fluctuations exhibit long-range
correlations with the power of the spectrum decaying as
$1/f^{\beta}$, with $\beta \sim 1$. Up until recently these
observations were considered a nuisance in the majority of
neuroimaging studies and disregarded as unwanted noise, despite
the fact that they are the baseline against which other
task-related conditions are usually compared.

The notion of a specific network of brain regions active in rest
states was reinforced by the observation of a consistent pattern
of deactivations seen across many goal-oriented tasks
\cite{shulman}. This observation coupled, with studies of cerebral
blood flow led Raichle and colleagues \cite{raichle_01} to propose
a theory for the so called brain ``default mode networks''. This
view sees BOLD signal decreases during cognitive tasks as one way
to identify how the brain is active during rest. In other words,
what part of the brain was more active during rest is inferred by
identifying what is being deactivated during a given task.

One simple way used to study this network is to look at the linear
correlations between the time series of BOLD activity of different
regions of the brain \cite{fox_05}. Figure 6 shows a typical
result from experiments in with the subjects were ask to track the
height of a moving bar varying in time during fMRI data
collection~\cite{baliki2}. The depicted correlation maps were
constructed by first extracting time series for the seeds (small
green circles in Fig. 6, obtaining averaging a cube of 3x3x3
voxels) and then computing its correlation coefficient with the
time series of all the other brain voxels. This is equivalent to
the correlation map shown previously for the Ising model (Fig. 5
top left panel).

\begin{figure}
\centering
\includegraphics[width=0.85 \textwidth]{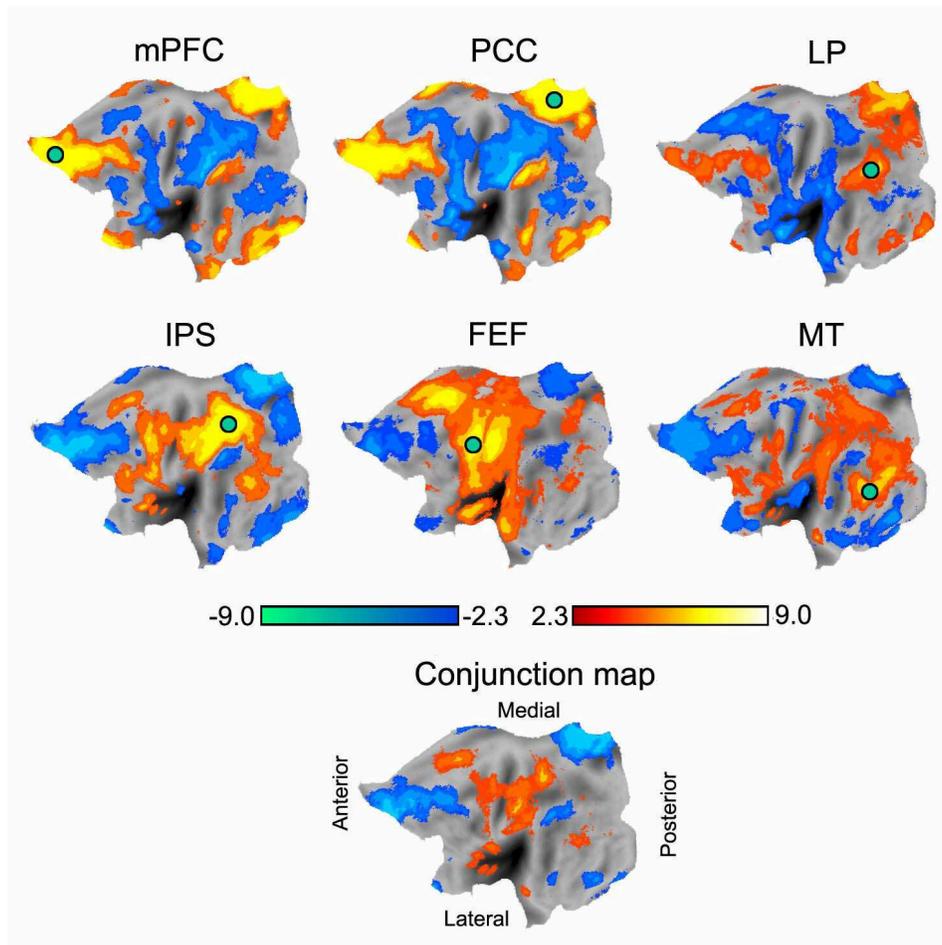}
\caption{Typical balanced correlated-anticorrelated spatial
domains of brain fMRI recorded from human volunteers during a
simple attention task (replotted from Baliki et al.
\cite{baliki2}. These patterns are typical of healthy individuals,
where the total area covered by positive correlations is
approximately the same as that covered by negatively correlations.
The data shows averaged z-score maps (for a group of 15
volunteers) showing regions with significant correlations with the
six seed regions (small circles).  The results shown correspond to
three task-negative seed regions: mPFC, PCC, and LP, as well as to
three task-positive seed regions: IPS, FEF, and MT. Colors
indicate regions with positive correlations (red-yellow) and
negative correlations (blue-green); both have z-scores $>$ 2.3
(p$<$0.01). The group z-score conjunction map below shows voxels
significantly correlated or anti-correlated with at least five of
the six seed regions.}
\end{figure}

Figure 6 shows correlation maps associated with six predefined
seed regions. Based on previous results, these seeds are known to
be sensitive to the task being conducted. Three regions, referred
to as task-positive regions, exhibit activity increases during the
task, and three regions, referred to as task-negative regions,
exhibit activity decreases (de-activation) during the attention
task~\cite{fox_05,corbetta}. Task-positive regions were centered
in the intraparietal sulcus (IPS), the frontal eye field region
(FEF), and the middle temporal region (MT). Task-negative regions
were centered in the medial prefrontal (mPFC), posterior
cingulate/precuneus (PCC), and lateral parietal cortex (LP).

The correlation maps of Figure 6 summarize the functional
co-activation between a given seed region and the rest of the
cortex. These maps replicate very closely the ones described at
rest~\cite{fox_05,mason,raichle_01,raichle_06a}, since it is known
that in minimally demanding tasks brain functional connectivity
approximates the functional connectivity seen during
rest~\cite{fox_05,greicius_03,greicius_04}. It displays  brain
regions that are positively correlated (red-yellow colors) and
regions that are negatively correlated (blue-green colors) with
any of the chosen seeds. An important experimental finding was
that the ratio between the area covered by positive correlations
and those with negative correlations was always very close to one
\cite{baliki2} (see Fig. 5 and Fig. 6). This was consistently
found in all healthy volunteers analyzed up to now. However, the
same analysis carried out in patients that have suffered chronic
pain for many years, revealed a ratio up to forty times larger
\cite{baliki2}. This suggests a healthy dynamic balance of the
resting state network, which deserves to be explored further.

The brain is clearly not a lattice and the connectivity is not
homogeneous. Moreover the ``small world'' features revealed by
fMRI described earlier are also found in the anatomical
connectivity~\cite{Sporns2004b}. Thus, finding in any given
complex spatiotemporal patterns what is due to the dynamics and
what is induced by the underlying structure is still a difficult
problem.

In an attempt to gain insight into the brain resting state
fluctuations, Honey et al.~\cite{honey}, simulated the cerebral
cortex using neuronal dynamics under the real structural
connections given by known large scale connectivity. According
with their results, coupled excitable elements embedded in this
type of anatomical architecture, favors the emergence of
spatio-temporal patterns such as those observed in the brain at
resting condition. For instance, they found that the functional
connectivity seen in the BOLD signal are present at low frequency
as a result of fluctuations in the aggregate number of transients
couplings and decoupling occurring at a more rapid scale ($\approx
10Hz$). At the slow time scale they identify two major
anti-correlated functional clusters which, in their
interpretation, are coordinated via anatomical connection
patterns. Nevertheless, the results shown in Fig. 5 suggest that
these anticorrelated clusters can be originated solely by the
critical dynamics.

The possibility which we favor is that the correlations seen
during resting state are very similar to those described for the
Ising model at the critical temperature. Of course, this
similarity is not in the details, but in the fundamental aspects
of the dynamics. In this view, spins are represented by entire
regions of coherent neuronal groups, say for instance any of the
seeds we choose in Figure 6. Thus, at each moment in time, each
cortical region competes or cooperates according with the
connectivity and the dynamics at that moment. The experimental
observation that at any given time positive and negative
correlations are equal is awaiting to be explained, and its
implications for disease further explored. We claim that the brain
is always near criticality such that the spatiotemporal patterns
illustrated above should be scale-invariant, and some other
temporal variables describing its evolution power law distributed.
If that is the case, then the resting state dynamical equivalent
is criticality as in other extended non-linear systems near the
edge of a phase transition.

\subsection{Epileptic seizures as brain quakes?}
In a recent paper, Osorio et al. \cite{Osorio2008} shows an
interesting analysis of the temporal organization of epileptic
seizures. They studied very large catalogues of seismic activity
and epileptic seizures with special attention to the statistical
distribution of event sizes, and waiting time between these
events. Their analysis reveals an striking analogy between the
dynamics of seizures and well known power laws governing
earthquakes such as the Gutemberg-Ritcher and Omori laws.

Some counterintuitive conclusions are worth to mention, already
noted for earthquakes \cite{bakbook}, such as the meaningless use
of intensity and duration to characterize a given seizure. This is
analogous to the scale invariance noticed already in the analysis
of earthquakes. In earthquakes, (as now seems in seizures) it is
known that to establish the probability of any event one must
specify a time window, a spatial grid size, and a given intensity
of that event.

Osorio et al.'s approach also elegantly answers the classical
question of why a seizure stops.  An earthquake stops
spontaneously whenever it has released the excess energy
accumulated. In geology terminology an earthquake ``starts without
knowing how big is going to be or how long it is going to last''.
Neuronal avalanches, according to Plenz' work described earlier,
also obey the same laws. According to the findings of Osorio et
al. the mechanism by which seizures stop is related with the same
critical process that triggers them. The authors comment that
``scale invariance in seizures may be conceptualized as the
hallmark of certain complex systems (the brain in this case) in
which, at or near the critical point, its component elements
(neurons) are correlated over all existing spatial (minicolum,
column, macrocolum, etc.) and temporal scales (microseconds,
seconds, tens of seconds, etc.)''

The similarities uncovered by Osorio et al. suggest that the
researchers' intuition regarding the statistical laws governing
epileptic seizures need to be adjusted accordingly.

\subsection{Senses are critical}
Of course brains are useful to escape from predators,  move
around, choose a mate or find food, and in these respects the
sensory apparatus is critical for any animal survival. Recent
results indicate that senses are also critical in the
thermodynamic sense of the word. Consider first the fact that the
density distribution of the various form of energy around us is
clearly inhomogeneous, at any level of biological reality, from
the sound loudness any animal have to adapt, to the amount of rain
a vegetable have to take advantage. From the extreme darkness of
a deep cave to the brightest flash of light there are several
order of magnitude changes; nevertheless our sensory apparatus is
able to inform the brain of such changes.
\begin{figure}
\centering
\includegraphics[width=0.8 \textwidth]{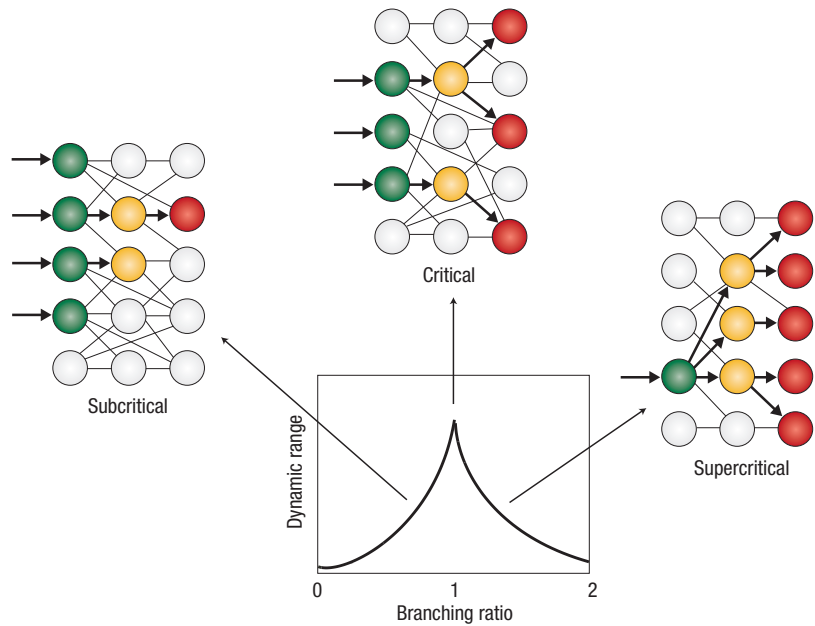}
\caption{Sensory networks constructed with branching ratios close
to one maintain, on the average, the input activity (green,
followed by yellow and red), thus optimizing the dynamic range. In
supercritical networks, however, activity explodes, while in
subcritical ones are unable to sustain any input pattern. Redrawn
from \cite{chialvo2006}.}
\end{figure}

It is well known that isolated neurons are unable to do that
because of their limited dynamic range, which spans only a single
order of magnitude. This is the oldest unsolved problem in the
field of psychophysics, tackled very recently by Kinouchi and
Copelli \cite{kinouchi} by showing that the dynamics emerging from
the \emph{interaction of coupled excitable elements}, is the key
to solving the problem. Their results show that a network of
excitable elements set precisely at the edge of a phase transition
- or, at criticality - can be both, extremely sensitive to small
perturbations and still able to detect large inputs without
saturation. This is generic for any network regardless of the
neurons' individual sophistication. The key aspect in the model is
a local parameter controlling the amplification of any initial
firing activity. Whenever the average amplification is very small
activity dies out, as it can be seen in the cartoon of Figure 7.
In this case the system is subcritical and not sensitive to small
inputs. On the other hand, choosing an amplification very large
one sets up the conditions for a supercritical reaction in which
for any - even very small - inputs the entire network fires. It is
only in between these two extremes that the networks have the
largest dynamic range. Thus, amplification around unity, i.e., at
criticality, seems to be the optimum condition for detecting large
energy changes as an animal encounters in the real world
\cite{chialvo2006}. Of course, in a critical world energy is
dissipated as a fractal in space and time with the characteristic
highly inhomogeneous fluctuations. As long as the world around is
critical, it seems that the  evolving organisms embedded in it
have no better choice than to be critical as well.

\section{Outlook}

The study of collective phenomena is at the center of statistical
physics. It is not surprising then, to see the recent outburst of
physics and mathematics publications studying this type of
phenomena in the context of computer science as well as social and
economic settings. While in all these fields there is a clear
transfer of methods and ideas from statistical physics, identical
flow has yet to start in brain science.

This lack of communication is even more intriguing if one
considers that most see brain science precisely as the study of
collective patterns of neuronal activity. Nonetheless, this
acknowledgment had not yet been translated into useful approaches.
To the contrary, the literature contains numerous old and new
promises to understand brain function by way of very large (and in
some cases very detailed) numerical simulations of millions of
neurons, completely orphan of considerations related with the
statistical physics of collective phenomena in large systems.

To make the point above relevance of these ideas, let us recall
once more the results presented in Figure 5 and the connectivity
between the sites revealed in Figure 4. As it was discussed, on
one side the connectivity says that the system is a lattice with
only four nearest neighbors, but the correlations reveal a network
with scale-free degree distribution. It seems to be a gross
contradiction, but the apparent divorce between the patterns
dictated by the coupling equations and those found by the analysis
of the spatial correlations will not surprise those already
familiar with emergent phenomena at the critical state. Again,
lets remind ourselves that the divorce between ``anatomy'' (i.e.,
the coupling) and dynamics disappears both in the supercritical
and subcritical state (as correlations vanish). Now, let suppose
that the time series data in Figure 5 were to be from a typical
brain experiment. Classical approaches of brain connectivity,
based either in the analysis of correlations (i.e., so-called
``functional'' connectivity or in the anatomical (i.e.,
``effective'') connections could never reach to the right
conclusion and solve the puzzle. It is only by knowing about the
features of critical phenomena that the apparent puzzle can be
solved. As far as we know, there is no report in the literature
suggesting changes in the character of the functional connectivity
due to the dynamics at the critical point as we suggest here.

In summary, according with the proposal reviewed here, several
relevant aspects of brain dynamics \emph{can} be only understood
using the theoretical framework as for any nonequilibrium
thermodynamic system  near the critical point of a second order
phase transition. That include the understanding of neuronal
dynamics at small scale, the cooperative-competitive equilibrium
seen at rest in the healthy cortex, the burst of brain quakes
during seizures and the optimization of the dynamic range at the
sensory periphery. We have mentioned but left out the discussion
of behavior, which understanding we submit should also benefit
from this approach.

Some of the ideas here are novel, but the motivation is not, since
Ashby was probably the first to indicate how fundamental is to
understand the way self-organization shapes brain function
\cite{Ashby}. Nevertheless, these views are gaining momentum, and
is refreshing to read recent reviews
\cite{Werner2006,werner2006b,werner2008} advocating the further
study of phase transitions, metastability and criticality in
cognitive models and experiments. This enlightening perspective is
even more meaningful coming from those that first introduced
information theory to the study of sensation in neuroscience,...
forty three years ago \cite{werner65}.

\begin{theacknowledgments}
Work supported by NIH NINDS of USA. Thanks to Drs. D. Plenz and
J.P. Segundo for stimulating discussions, and to E. Parks for
proofreading the manuscript. PB and DF are researchers supported
by CONICET, Argentina.
\end{theacknowledgments}

\end{document}